\documentclass[a4paper,aps,prd,10pt,preprintnumbers,showpacs,twocolumn,superscriptaddress,nofootinbib,amsmath,amssymb,gensymb,amsfonts]{revtex4-1}
\usepackage{graphicx}
\usepackage{cmap}
\usepackage[utf8]{inputenc}
\usepackage[T1]{fontenc}
\usepackage{nicefrac}
\usepackage{bm,amsmath,amssymb,amsfonts,gensymb,bbold}
\usepackage[colorlinks=true,linktocpage=true,linkcolor=blue,citecolor=blue,allcolors=blue]{hyperref}
\usepackage{amsmath}
\usepackage{makecell}
\usepackage{soul}
\usepackage{placeins}
\usepackage{epstopdf}
\graphicspath{{../}}
\def\imo{i}

\begin{document}
\title{Long-lived Quasinormal modes around regular black holes and wormholes in Covariant Effective Quantum Gravity}
\author{B. C. Lütfüoğlu}
\email{bekir.lutfuoglu@uhk.cz}
\affiliation{Department of Physics, Faculty of Science, University of Hradec Kralove, \\
Rokitanskeho 62/26, Hradec Kralove, 500 03, Czech Republic. }

\begin{abstract}
We study quasinormal modes of massive scalar and massless Dirac fields in the background of regular black holes and traversable wormholes arising in Covariant Effective Quantum Gravity. Using both the Jeffreys-Wentzel-Kramers-Brillouin approximation and time-domain integration, we analyze the impact of quantum corrections on the quasinormal spectra and late-time behavior of perturbations. Our results reveal the existence of slowly decaying, oscillatory tails and quasi-resonant modes in the scalar sector, particularly in the high-mass regime. In the fermionic case, the damping rate increases with the quantum correction parameter \( \xi \), while the oscillation frequency decreases. We also observe pronounced echo-like structures in the time-domain profiles near the black hole–wormhole threshold. These findings provide insight into the dynamics of perturbations in quantum-corrected spacetimes and offer potential signatures for distinguishing black holes from wormholes in future gravitational wave observations.
\end{abstract}

\maketitle
\section{Introduction}

The study of quasinormal modes (QNMs) of black holes \cite{Kokkotas:1999bd,Nollert:1999ji,Konoplya:2011qq,Bolokhov:2025uxz} provides a crucial link between gravitational wave observations \cite{LIGOScientific:2017vwq, LIGOScientific:2020zkf, LISA:2022kgy, NANOGrav:2023gor}, observations in electromagnetic spectrum \cite{EventHorizonTelescope:2019dse,EventHorizonTelescope:2019ggy,Goddi:2016qax} and the fundamental properties of compact objects, such as mass, charge, and spin. In recent years, growing attention has been paid to the effects of quantum corrections on the geometry of black holes, motivated by attempts to resolve singularities and incorporate semiclassical gravity effects. Quantum-corrected black holes, emerging from various approaches such as loop quantum gravity, asymptotically safe gravity, or non-perturbative effective field theory, exhibit modified spacetime structures near the event horizon. As a result, a substantial number of studies have focused on the QNMs for various models of quantum-corrected black holes \cite{Bolokhov:2023ruj,Gong:2023ghh,Fu:2023drp,Malik:2024tuf,Moreira:2023cxy,Baruah:2023rhd,Bolokhov:2023bwm,Chen:2022ngd,Cruz:2020emz,Liu:2012ee,Konoplya:2001ji,Konoplya:2023ahd,Piedra:2009pf,Anacleto:2021qoe,Campos:2021sff,Hamil:2024ppj,Skvortsova:2024atk,Zinhailo:2024kbq,Zinhailo:2018ska,Malik:2024nhy,Malik:2024elk,Lin:2024ubg,Campos:2021sff,Anacleto:2021qoe}. Notably, Konoplya et al. \cite{Konoplya:2022pbc} recently demonstrated that these quantum modifications can leave discernible imprints on the QNM spectra, particularly in the high-overtone regime or during the early stages of the ringdown phase.

A particularly interesting approach to constructing quantum-corrected black hole spacetimes was proposed in \cite{Zhang:2024ney}, where the authors employed Hamiltonian constraints inspired by loop quantum gravity \cite{Thiemann:2007zz,Ashtekar:2004eh,Zhang:2024khj}. This method yields a quantum-corrected solution that describes a regular black hole, fulfilling long-standing expectations from quantum gravity. Remarkably, the same family of solutions also admits traversable wormhole geometries, depending on the value of the quantum correction parameter $\xi$. Another notable feature of these quantum-corrected black holes is the absence of Cauchy horizons, which enhances the stability of the model compared to earlier proposals for regular black holes developed in \cite{Zhang:2024khj,Konoplya:2024lch}.

The idea of transitions between regular black holes and wormholes has attracted considerable interest and has been explored in numerous works \cite{Simpson:2018tsi,Casadio:2001jg,Bronnikov:2002rn,Bronnikov:2021uta,Alencar:2025jvl,Churilova:2019cyt}. Unlike the quantum-corrected solutions in \cite{Zhang:2024ney}, these models do not emerge as a single scenario within a well-motivated theory of gravity. Recently, the QNMs of massless bosonic fields in such quantum-corrected spacetimes were studied in \cite{Konoplya:2025hgp}. However, to date, there have been no corresponding studies for massive bosonic fields or even massless fermionic fields. This gap is particularly noteworthy given the substantial body of work devoted to Dirac field perturbations in classical black hole backgrounds~\cite{Bolokhov:2024ixe,Gonzalez:2014voa,Jing:2005uy,Konoplya:2017tvu,Rosa:2011my,Kanti:2006ua,Al-Badawi:2023xig,Chen:2005rm,Saleh:2016pke,Varghese:2010qv,Sebastian:2014dka,Wu:2004vb,Zhang:2005zs,Wang:2017fie,Konoplya:2007zx,Jing:2003wq,Zhou:2003ts}, largely motivated by the role of neutrinos in astrophysical and cosmological black hole environments.

The study of QNMs of massive scalar fields unveils a wide range of intriguing physical phenomena, expanding upon the well-established framework for massless fields. Introducing a mass term in the field equations enriches the dynamics significantly and leads to qualitatively new features in both the quasinormal spectrum and the late-time evolution of perturbations \cite{Konoplya:2004wg,Konoplya:2017tvu,Zhidenko:2006rs,Ohashi:2004wr,Zhang:2018jgj,Aragon:2020teq,Ponglertsakul:2020ufm,Gonzalez:2022upu,Burikham:2017gdm}. 

One of the most remarkable aspects is the appearance of arbitrarily long-lived modes, often referred to as \textit{quasi-resonances} \cite{Ohashi:2004wr,Konoplya:2004wg}. These modes are known to emerge for various spin fields \cite{Fernandes:2021qvr,Konoplya:2017tvu,Percival:2020skc} and have been identified in the spectra of numerous compact objects, such as black holes \cite{Zhidenko:2006rs,Zinhailo:2018ska,Bolokhov:2023bwm,Skvortsova:2024eqi} and wormholes \cite{Churilova:2019qph}. However, the existence of quasi-resonances is not universal: in certain spacetimes or field configurations, such modes are absent \cite{Zinhailo:2024jzt,Lutfuoglu:2025hjy}, underscoring the need for careful, case-specific analysis, including in quasiblack hole geometries where long-lived modes have also been reported~\cite{Guo:2021bcw}.

From a phenomenological perspective, the motivation to study massive field perturbations stems from a variety of physical models. In braneworld scenarios, for instance, mass terms can arise effectively due to the projection of higher-dimensional bulk effects onto the brane, thereby influencing the dynamics of brane-localized fields \cite{Seahra:2004fg}. Similarly, in massive gravity theories, massive modes may carry implications for gravitational wave physics, particularly in the ultra-low-frequency range explored by pulsar timing array experiments \cite{Konoplya:2023fmh, NANOGrav:2023hvm}. 

Moreover, mass terms are not restricted to fundamental properties of fields but can emerge effectively. For example, a massless scalar propagating in a black hole spacetime immersed in an external magnetic field acquires an effective mass, significantly altering its quasinormal mode spectrum \cite{Konoplya:2007yy,Konoplya:2008hj,Wu:2015fwa,Davlataliev:2024mjl,Kokkotas:2010zd,Chen:2011jgd}.

Another key motivation lies in the distinct behavior of massive fields at late times. Unlike their massless counterparts, which exhibit a characteristic power-law decay, massive fields decay via oscillatory tails \cite{Jing:2004zb,Koyama:2001qw,Moderski:2001tk,Rogatko:2007zz,Koyama:2001ee,Koyama:2000hj,Gibbons:2008gg,Gibbons:2008rs,Konoplya:2006gq}. This difference not only has theoretical importance but also contributes to potential observational signatures in gravitational wave astronomy.

%{\color{red} I think it would be better to give the aim near the end of the Introduction. I also added several sentences.}
%{\color{blue}In this work, we aim to complement previous studies by analyzing the quasinormal spectrum of massless Dirac and massive scalar perturbations of quantum-corrected black holes and traversable wormholes arising in the framework of Covariant Effective Quantum Gravity.}
In this work, we aim to complement previous studies by analyzing the quasinormal spectrum of massless Dirac and massive scalar perturbations of quantum-corrected black holes and traversable wormholes arising in the framework of Covariant Effective Quantum Gravity. To this end, we employ both the Jeffreys-Wentzel-Kramers-Brillouin (JWKB) approximation and time-domain integration to extract the characteristic frequencies and investigate the impact of quantum corrections on the dynamics of perturbations. Our analysis focuses in particular on the emergence of long-lived modes, quasi-resonances, and echo-like structures near the black hole–wormhole transition, which may serve as potential observational signatures of the underlying quantum-modified geometry.

The paper is organized as follows. In Sec.~\ref{sec:metric}, we briefly review the quantum-corrected black hole and wormhole solution derived from the Hamiltonian constraints approach within Covariant Effective Quantum Gravity. Sec.~\ref{sec:perturbations} is devoted to the derivation of the perturbation equations and the corresponding effective potentials for massive scalar and massless Dirac fields. In Sec.~\ref{sec:methods}, we outline the numerical and semi-analytical techniques used to extract the quasinormal frequencies, including the JWKB method and time-domain integration. Sec.~\ref{sec:qnms} presents a detailed analysis of the quasinormal spectra, discussing the effects of mass, coupling strength, and the black hole–wormhole transition. Finally, in Sec.~\ref{sec:conclusions}, we summarize our results and discuss possible directions for future research.

\section{Regular black hole and wormhole metric}\label{sec:metric}

In Refs.~\cite{Zhang:2024khj, Zhang:2024ney}, the authors constructed a modified Hamiltonian constraint designed to encapsulate quantum gravitational effects while maintaining full diffeomorphism invariance, notably without imposing any particular gauge fixing. By appropriately selecting the functional form of the free parameters in the theory, they succeeded in deriving a class of solutions that describe either regular black holes or traversable wormholes, with the specific outcome governed by the values of the quantum correction parameters \cite{Zhang:2024ney}. 

A key advantage of this framework over earlier quantum-corrected black hole models lies in the absence of Cauchy horizons within the solution. Since Cauchy horizons are typically associated with instabilities under linear perturbations, their elimination in this setting may point to a more robust and physically viable description of quantum-corrected compact objects.

The quantum-corrected black hole metric introduced in~\cite{Zhang:2024ney} is described by the following line element:
\begin{equation}\label{eq:metric}
ds^2 = -f(r)\, dt^2 + \frac{1}{f(r)\mu(r)}\, dr^2 + r^2\, d\Omega^2,
\end{equation}
where \( d\Omega^2 = d\theta^2 + \sin^2\theta\, d\phi^2 \) denotes the metric on a unit two-sphere. The metric functions are defined as
\begin{align}
f(r) &= 1 - (-1)^n \frac{r^2}{\xi^2} \arcsin\left( \frac{2M \xi^2}{r^3} \right) - \frac{n \pi r^2}{\xi^2}, \\
\mu(r) &= 1 - \frac{4 M^2 \xi^4}{r^6},
\end{align}
where \( \xi \) is a parameter encoding the quantum corrections, \( M \) denotes the total mass of the configuration, and \( n \in \mathbb{Z} \) is an arbitrary integer indexing distinct branches of the solution.

For the purposes of this study, we focus on the case \( n = 0 \), corresponding to asymptotically flat spacetimes. In this setting, the nature of the solution crucially depends on the ratio \( \xi/M \). When \( \xi/M < \pi^{3/2}/\sqrt{2} \approx 3.937 \), the geometry describes a regular black hole in which the central singularity is effectively replaced by a smooth geometry with a minimal surface, or “throat,” located at
\[
r_m = \sqrt[3]{2}\, M^{1/6} \xi^{2/3}.
\]
However, for larger values of \( \xi \), the black hole horizon ceases to exist, and the solution describes a traversable wormhole connecting two asymptotically flat regions.

The observational properties of these geometries have been partially explored. In particular, the shadow cast by the black hole and wormhole configurations was investigated in~\cite{Liu:2024iec}, while gravitational lensing effects were discussed in~\cite{Paul:2025wen}.

To facilitate the analysis of the wormhole sector of the solution, it is convenient to reparametrize the metric using quasiglobal coordinates. Introducing a new radial coordinate \( x \in (-\infty, \infty) \) defined via
\begin{equation}\label{eq:rtox}
R(x) = \sqrt{x^2 + r_m^2},
\end{equation}
where \( x = 0 \) corresponds to the wormhole throat and \( x \to \pm \infty \) denotes the asymptotic regions of two separate universes, the metric~\eqref{eq:metric} takes the form
\begin{equation}\label{eq:metric_quasiglobal}
ds^2 = -A(x)\, dt^2 + B(x)\, dx^2 + R^2(x)\, d\Omega^2,
\end{equation}
with the metric functions given by
\begin{align}
A(x) &= f(R(x)), \\
B(x) &= \frac{x^2}{(x^2 + r_m^2)\, f(R(x))\, \mu(R(x))}.
\end{align}

This coordinate system provides a clearer view of the wormhole geometry, especially for analyzing the causal structure and studying wave propagation. Additional optical signatures and observational implications of these geometries have also been addressed in~\cite{Liu:2024iec,Paul:2025wen}.

\section{Perturbation equations and effective potential}\label{sec:perturbations}

The dynamics of a massive scalar and massles Dirac fields in curved spacetime are governed by the following general covariant equations
\begin{subequations}\label{coveqs}
\begin{eqnarray}\label{KGg}
\frac{1}{\sqrt{-g}} \partial_\mu \left( \sqrt{-g} g^{\mu \nu} \partial_\nu \Phi \right) &=& \mu^2 \Phi, 
\\ \label{covdirac}
\gamma^{\alpha} \left( \frac{\partial}{\partial x^{\alpha}} - \Gamma_{\alpha} \right) \Upsilon&=&0,
\end{eqnarray}
\end{subequations}
Here $\gamma^{\alpha}$ are the curved-spacetime Dirac matrices, and $\Gamma_\alpha$ are the associated spin connections.

By applying a separation of variables procedure to these equations in the background metric (\ref{eq:metric_quasiglobal}), one can reduce the problem to a single Schrödinger-like wave equation for each type of perturbation. This master equation describes the evolution of perturbation functions $\Psi$ in terms of the tortoise coordinate $r_*$ and an effective potential $V(r_*)$ that depends on the background geometry:
\begin{equation}
\left(\frac{\partial^2}{\partial t^2} - \frac{\partial^2}{\partial r_*^2}\right)\Psi + V(r_*) \Psi = 0, 
\end{equation}
where the tortoise coordinate $r_*$ is defined via $dx/dr_* = \sqrt{A/B}$, ensuring that the wave equation assumes a canonical form.

The explicit forms of the effective potentials for the different fields are given by:
\begin{equation}
V(x) = \frac{1}{2} \left(\frac{A}{B}\right)' \frac{R'}{R} + \frac{A}{B} \frac{R''}{R} + \frac{A}{R^2} \ell(\ell+1) + \frac{A}{R^2}  \mu^2, 
\end{equation}
For the Dirac field ($s=1/2$) one has two isospectral potentials,
\begin{equation}
V_{\pm}(r) = W^2\pm\frac{dW}{dr_*}, \quad W \equiv 
\left(\ell+\frac{1}{2}\right)\frac{\sqrt{1-\frac{r^2 \sin^{-1}\left(\frac{2 M \xi^2}{r^3}\right)}{\xi ^2}}}{r}.
\end{equation}
The isospectral wave functions can be transformed one into another by the Darboux transformation,
\begin{equation}\label{psi}
\Psi_{+}\propto \left(W+\dfrac{d}{dr_*}\right) \Psi_{-},
\end{equation}
so that it is sufficient to calculate QNMs for only one of the effective potentials. We will do that for $V_{+}(r)$ because the JWKB method works better in this case.

Transforming to the frequency domain using the ansatz $$\Psi(t, r_*) = e^{-i \omega t} \Psi(r_*),$$ we obtain the time-independent form of the wave equation:
\begin{equation}\label{eq:master_eq_freq}
\frac{d^2}{dr_*^2} \Psi + \left[\omega^2 - V(r_*)\right] \Psi = 0.
\end{equation}

\begin{figure}
\resizebox{\linewidth}{!}{\includegraphics{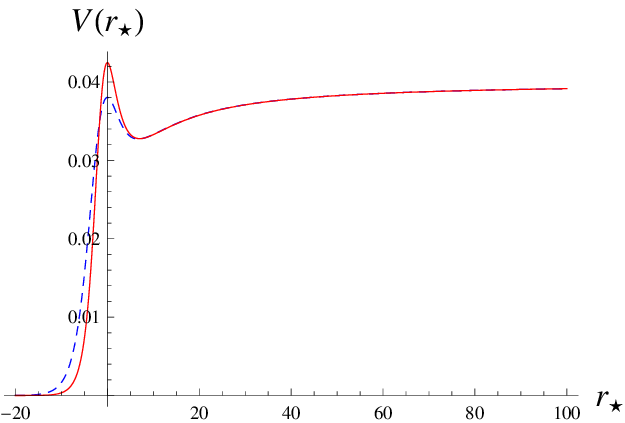}}
\caption{Effective potential as a function of the tortoise coordinate $r^{*}$ for $\ell=0$ scalar field perturbations; $M=1$, $\xi =0.1$ (blue, dashed line) and $\xi =2.5$ (red, solid line).}\label{fig:scalarpot}
\end{figure}

\begin{figure}
\resizebox{\linewidth}{!}{\includegraphics{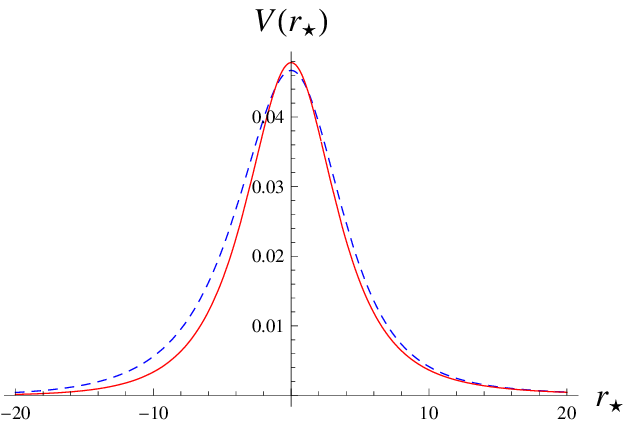}}
\caption{Effective potential as a function of the tortoise coordinate $r^{*}$ for $\ell=1/2$ Dirac field perturbations; $M=1$, $\xi =0.1$ (blue, dashed line) and $\xi =2.5$ (red, solid line).}\label{fig:Diracpot}
\end{figure}

\begin{figure*}                            
\resizebox{\linewidth}{!}{\includegraphics{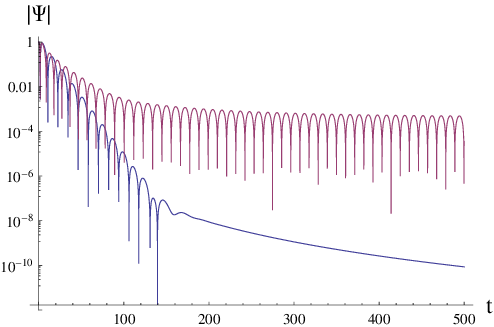}\includegraphics{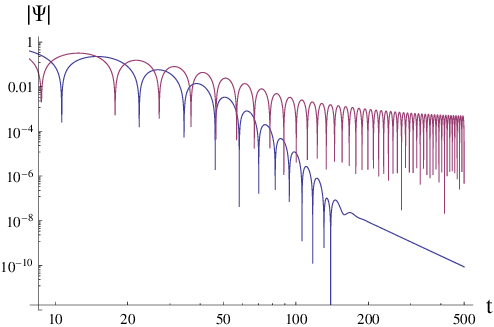}}          
\caption{Semi-logarithmic (left) and logarithmic (right) time-domain profiles for regular BH perturbations: $\ell=1$, $\xi =3.5$, $M=1$, $\mu=0$ (blue) and $\mu=0.3$ (red). The Prony method gives $\omega= 0.263327 - 0.118669 i$ (for $\mu=0$) and  $\omega= 0.290274 - 0.037524 i$ (for $\mu=0.3$). }\label{fig:TD1}
\end{figure*}

\begin{figure}                            
\resizebox{\linewidth}{!}{\includegraphics{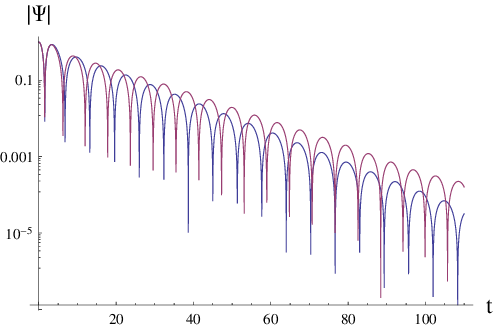}}          
\caption{Time-domain profiles for regular BH perturbations: $\ell=2$, $\xi =0.1$, $M=1$, $\mu=0.2$ (blue) and $\mu=0.4$ (red). The fundamental ($n=0$) QNMs given by the WKB method for these values of the parameters are: $\omega= 0.4963 - 0.0924 i $ (for  $\mu=0.2$) and $\omega= 0.5350 - 0.0787 i$ (for  $\mu=0.4$ ), while the corresponding WKB values are $\omega= 0.496326-0.092389 i$ and $\omega = 0.535102-0.078675 i$.}\label{fig:TD2}
\end{figure}

Typical forms of the effective potentials are shown in Figs. \ref{fig:scalarpot} and \ref{fig:Diracpot}, where one can see that they are positive definite and, in the case of a massive scalar field, asymptote $\mu^2$ at spatial infinity. Therefore, the perturbations under consideration must be stable, not allowing for unboundedly growing modes.

\section{Methods}\label{sec:methods}

Quasinormal frequencies of black holes and wormholes $\omega$ are  oscillation frequencies satisfying the same boundary conditions in terms of the tortoise coordinate 
\begin{equation}\label{boundaryconditions1}
\Psi(r_*\to \infty) \propto e^{\imo \omega r_*}, 
\end{equation}
\begin{equation}\label{boundaryconditions2}
\Psi(r_*\to - \infty) \propto e^{- \imo \omega r_*}, 
\end{equation}
which correspond to purely incoming waves at the event horizon (for a black hole) \cite{Kokkotas:1999bd,Nollert:1999ji} or at spatial infinity in the second universe (for a wormhole \cite{Churilova:2021tgn,Malik:2024wvs}) ($r_*\to-\infty$), and purely outgoing wave at spatial infinity ($r_*\to\infty$) in both cases, respectively.

Here, we will use two alternative methods for finding QNMs: JWKB method and time-domain integration.

\subsection{JWKB method}

Among the various analytical and semi-analytical techniques developed to compute QNM spectra, the JWKB approximation has proven to be particularly versatile and widely adopted. Its popularity stems from a favorable combination of computational efficiency, ease of automation, and often sufficiently high accuracy across a broad range of applications (see, e.g., \cite{Barrau:2019swg,Stashko:2024wuq,Konoplya:2001ji,Zinhailo:2019rwd,Xiong:2021cth,Skvortsova:2024atk,Konoplya:2006ar,Skvortsova:2024wly,Hamil:2024nrv,Liu:2024wch,Konoplya:2005sy,Kodama:2009bf}).

The core idea behind the JWKB method is to approximate the effective potential near its maximum by a Taylor series expansion, allowing one to match the asymptotic solutions of the wave equation across two classical turning points. These asymptotic solutions are constrained by the boundary conditions characteristic of QNMs, namely, purely outgoing waves at spatial infinity and purely ingoing waves at the event horizon. Consequently, the JWKB method is particularly well suited to systems where the effective potential forms a single-peak barrier—a condition often met in black hole perturbation theory.

In the specific context of black holes surrounded by massive fields and influenced by quantum corrections through the parameter~$\xi$, the effective potential has three turning points. Therefore, the JWKB approach yields sufficiently accurate results for the quasinormal spectrum only when the mass term is sufficiently small. Exceptions are the cases of asymptotically de Sitter spacetimes when the potential has a single peak again \cite{Fontana:2020syy,Dubinsky:2024hmn}.

The JWKB approximation provides a semi-analytic method for calculating QNMs by expanding the wave function near the peak of the effective potential. In its standard form, the JWKB formula at \( N \)th order takes the form~\cite{Iyer:1986np,Konoplya:2003ii}:
\begin{equation}
\frac{i Q_0}{\sqrt{2 Q_0''}} - \sum_{j=2}^{N} \Lambda_j = n + \frac{1}{2}, \quad n = 0,1,2,\dots,
\end{equation}
where \( Q(r) = \omega^2 - V(r) \), \( Q_0 \) is the value of \( Q(r) \) at its maximum, \( Q_0'' \) is the second derivative with respect to the tortoise coordinate at the peak, and \( \Lambda_j \) are higher-order JWKB correction terms.

Although higher-order JWKB methods (such as 6th or 13th order) improve accuracy, the convergence is not guaranteed, and in some cases the results oscillate with increasing order. To resolve this, one can apply Padé approximants to the JWKB expansion series. The Padé approximant is a rational function of the form
\begin{equation}
P_{m/n}(x) = \frac{a_0 + a_1 x + \dots + a_m x^m}{1 + b_1 x + \dots + b_n x^n},
\end{equation}
constructed to match the series expansion up to order \( m+n \). This resummation technique captures the analytic structure of the function better than a truncated Taylor expansion.

In the context of QNMs, Padé approximants are used to improve the convergence of the JWKB series for \( \omega \). Typically, the Padé approximant is applied to the square of the frequency \( \omega^2 \), written as:
\begin{equation}
\omega^2 = V_0 + \sum_{k=1}^{N} \Pi_k(\omega), 
\end{equation}
where \( \Pi_k \) are JWKB correction terms depending on \( \omega \), and the Padé approximant \( P^{N}_\ell(\omega^2) \) is constructed from this series.

It has been shown~\cite{Matyjasek:2017psv} that the use of Padé approximants significantly improves the accuracy and stability of the JWKB method, especially for low multipole numbers and overtone modes, where the standard JWKB expansion may fail to converge.

The systematic derivation of these correction terms has been accomplished at various orders: the second and third orders were developed in the seminal work of Iyer and Will \cite{Iyer:1986np}, fourth through sixth orders were presented in \cite{Konoplya:2003ii}, and extensions up to the thirteenth order were achieved in \cite{Matyjasek:2017psv}. These advancements have rendered the JWKB method not only accurate for fundamental modes but also reliable for higher overtones, particularly when used in conjunction with Padé resummation techniques, which further improve its convergence and applicability.

Thus, the JWKB method remains one of the most practical and widely applicable tools for computing QNMs, especially in scenarios where the effective potential retains the necessary single-barrier structure, as is the case for many quantum-corrected and alternative gravity black hole models.

\subsection{Time-domain integration}

To investigate the dynamical response of a black hole or wormhole to external perturbations, we analyze the time evolution of the associated wave function at a fixed radial position. This is achieved through numerical integration in the time domain, enabling a direct visualization of how perturbations propagate and decay over time. For this purpose, we adopt a finite-difference discretization approach introduced by Gundlach, Price, and Pullin~\cite{Gundlach:1993tp}, which has become a standard method in the study of black hole perturbations.  This method has been employed in a number of studies, demonstrating excellent agreement with other independent approaches~\cite{Skvortsova:2023zca,Konoplya:2017ymp,Konoplya:2020jgt,Konoplya:2013sba,Cuyubamba:2016cug}.

The discretization scheme is implemented on a two-dimensional numerical grid defined by double-null coordinates \( u \) and \( v \), where the wave function is evaluated at successive grid points. The update formula for the wave function at the grid point \( N = (u + \Delta, v + \Delta) \) is given by:
\begin{eqnarray}
\Psi\left(N\right) &=& \Psi\left(W\right) + \Psi\left(E\right) - \Psi\left(S\right) \nonumber \\
&-& \frac{\Delta^2}{8} V\left(S\right) \left[ \Psi\left(W\right) + \Psi\left(E\right) \right] + \mathcal{O}(\Delta^4),
\label{Discretization}
\end{eqnarray}
where the labels for the grid points are defined as:
\begin{align*}
N &\equiv (u + \Delta, v + \Delta), \quad &\text{(North)}, \\
W &\equiv (u + \Delta, v), \quad &\text{(West)}, \\
E &\equiv (u, v + \Delta), \quad &\text{(East)}, \\
S &\equiv (u, v). \quad &\text{(South)}.
\end{align*}
Here, \( \Delta \) is the step size in both null directions, and \( V(S) \) is the value of the effective potential evaluated at the grid point \( S \). The inclusion of higher-order terms proportional to \( \Delta^4 \) enhances the accuracy of the evolution scheme for sufficiently small step sizes.

To determine the quasinormal mode spectrum from the resulting time-domain profile, we employ the Prony method—a well-established technique for extracting characteristic frequencies from damped oscillatory signals. This method models the waveform as a sum of exponentially decaying sinusoidal components:
\begin{equation}
\Psi(t) \approx \sum_{i=1}^{p} C_i e^{-i\omega_i t},
\label{damping-exponents}
\end{equation}
where \( C_i \) are complex amplitudes, and \( \omega_i \in \mathbb{C} \) are the complex frequencies associated with each mode. The real part of \( \omega_i \) encodes the oscillation frequency, while its imaginary part governs the decay rate of the perturbation. 

By fitting this model to the portion of the signal corresponding to the dominant ringdown phase—typically just after the initial transient has subsided—we can isolate the fundamental quasinormal mode and, if desired, the first few overtones. This procedure provides valuable information about the underlying geometry and stability of the compact object, and plays a key role in gravitational wave astrophysics, where such modes serve as fingerprints of black hole spacetimes.

\begin{table}
\begin{tabular}{c c c c}
\hline
\hline
$\xi$ & WKB6Padé3 & WKB7Padé3 & rel. diff.  \\
\hline
$0.4$ & $0.112094-0.104590 i$ & $0.111934-0.104988 i$ & $0.280\%$\\
$0.8$ & $0.112512-0.104448 i$ & $0.110950-0.106956 i$ & $1.92\%$\\
$1.2$ & $0.111434-0.104703 i$ & $0.109896-0.105656 i$ & $1.18\%$\\
$1.6$ & $0.109263-0.107084 i$ & $0.109598-0.105002 i$ & $1.38\%$\\
$2.0$ & $0.111488-0.113003 i$ & $0.100096-0.110789 i$ & $7.31\%$\\
$2.4$ & $0.099489-0.112733 i$ & $0.092836-0.110997 i$ & $4.57\%$\\
$2.8$ & $0.093559-0.108021 i$ & $0.086980-0.108803 i$ & $4.64\%$\\
$3.2$ & $0.077644-0.109507 i$ & $0.077559-0.110179 i$ & $0.504\%$\\
$3.6$ & $0.060761-0.102910 i$ & $0.063945-0.111595 i$ & $7.74\%$\\
$3.7$ & $0.053750-0.101172 i$ & $0.059977-0.112398 i$ & $11.2\%$\\
\hline
\hline
\end{tabular}
\caption{QNMs of the $\ell=0$ test scalar field ($s=0$) for the quantum corrected black hole ($M=1$) calculated using the JWKB formula at different orders and Padé approximants. The deviation is given in per cents.}\label{table1}
\end{table}

\begin{table}
\begin{tabular}{c c c c}
\hline
\hline
$\xi$ & WKB6Padé3 & WKB7Padé3 & rel. diff.  \\
\hline
$0.4$ & $0.292932-0.097683 i$ & $0.292928-0.097680 i$ & $0.0016\%$\\
$0.8$ & $0.292839-0.097854 i$ & $0.292825-0.097952 i$ & $0.0319\%$\\
$1.2$ & $0.292328-0.098968 i$ & $0.292339-0.099019 i$ & $0.0169\%$\\
$1.6$ & $0.290821-0.101475 i$ & $0.290674-0.101383 i$ & $0.0565\%$\\
$2.0$ & $0.287720-0.105173 i$ & $0.287493-0.105082 i$ & $0.0796\%$\\
$2.4$ & $0.282938-0.109260 i$ & $0.282622-0.109559 i$ & $0.144\%$\\
$2.8$ & $0.276563-0.112843 i$ & $0.276341-0.112544 i$ & $0.125\%$\\
$3.2$ & $0.266443-0.115594 i$ & $0.268227-0.116371 i$ & $0.670\%$\\
$3.6$ & $0.258739-0.114735 i$ & $0.260490-0.118873 i$ & $1.59\%$\\
$3.7$ & $0.257311-0.114895 i$ & $0.258531-0.119353 i$ & $1.64\%$\\
\hline
\hline
\end{tabular}
\caption{QNMs of the $\ell=1$ test scalar field ($s=0$) for the quantum corrected black hole ($M=1$) calculated using the JWKB formula at different orders and Padé approximants. The deviation is given in per cents.}\label{table2}
\end{table}

\begin{table}
\begin{tabular}{c c c c}
\hline
\hline
$\xi$ & WKB6Padé3 & WKB7Padé3 & rel. diff. \\
\hline
$0.4$ & $0.483634-0.096773 i$ & $0.483635-0.096773 i$ & $0.00014\%$\\
$0.8$ & $0.483493-0.096986 i$ & $0.483493-0.096985 i$ & $0.00023\%$\\
$1.2$ & $0.482853-0.097874 i$ & $0.482884-0.097886 i$ & $0.00678\%$\\
$1.6$ & $0.481084-0.099957 i$ & $0.481107-0.100005 i$ & $0.0107\%$\\
$2.0$ & $0.477385-0.103342 i$ & $0.477377-0.103337 i$ & $0.00185\%$\\
$2.4$ & $0.471302-0.107381 i$ & $0.471308-0.107378 i$ & $0.00141\%$\\
$2.8$ & $0.463230-0.111150 i$ & $0.463146-0.111137 i$ & $0.0179\%$\\
$3.2$ & $0.453840-0.114076 i$ & $0.453805-0.114387 i$ & $0.0669\%$\\
$3.6$ & $0.443498-0.116086 i$ & $0.443488-0.116833 i$ & $0.163\%$\\
$3.7$ & $0.440816-0.116491 i$ & $0.440856-0.117313 i$ & $0.180\%$\\
\hline
\hline
\end{tabular}
\caption{QNMs of the $\ell=2$ test scalar field ($s=0$) for the quantum corrected black hole ($M=1$) calculated using the JWKB formula at different orders and Padé approximants. The deviation is given in per cents.}\label{table3}
\end{table}

\begin{table}
\begin{tabular}{c c c c}
\hline
\hline
$\xi$ & WKB6Padé3 & WKB7Padé3 & rel. diff. \\
\hline
$0.4$ & $0.182620-0.096599 i$ & $0.182449-0.096588 i$ & $0.0830\%$\\
$0.8$ & $0.182363-0.096980 i$ & $0.182449-0.096920 i$ & $0.0508\%$\\
$1.2$ & $0.182245-0.098302 i$ & $0.182260-0.097999 i$ & $0.147\%$\\
$1.6$ & $0.181440-0.100446 i$ & $0.181437-0.100177 i$ & $0.130\%$\\
$2.0$ & $0.175121-0.102115 i$ & $0.177956-0.103449 i$ & $1.55\%$\\
$2.4$ & $0.179180-0.098460 i$ & $0.173837-0.106105 i$ & $4.56\%$\\
$2.8$ & $0.174190-0.101934 i$ & $0.168507-0.106614 i$ & $3.65\%$\\
$3.2$ & $0.168447-0.102835 i$ & $0.162910-0.107309 i$ & $3.61\%$\\
$3.6$ & $0.162296-0.102449 i$ & $0.157083-0.107854 i$ & $3.91\%$\\
$3.7$ & $0.160737-0.102259 i$ & $0.155633-0.107901 i$ & $3.99\%$\\
\hline
\hline
\end{tabular}
\caption{QNMs of the $\ell=1/2$ test scalar field ($s=0$) for the quantum corrected black hole ($M=1$) calculated using the JWKB formula at different orders and Padé approximants. The deviation is given in per cents.}\label{table4}
\end{table}

\begin{table}
\begin{tabular}{c c c c}
\hline
\hline
$\xi$ & WKB6Padé3 & WKB7Padé3 & rel. diff. \\
\hline
$0.4$ & $0.380027-0.096413 i$ & $0.380031-0.096418 i$ & $0.0017\%$\\
$0.8$ & $0.379916-0.096609 i$ & $0.379919-0.096612 i$ & $0.0011\%$\\
$1.2$ & $0.379412-0.097417 i$ & $0.379404-0.097409 i$ & $0.0029\%$\\
$1.6$ & $0.377931-0.099295 i$ & $0.377890-0.099283 i$ & $0.0108\%$\\
$2.0$ & $0.374798-0.102122 i$ & $0.374784-0.102181 i$ & $0.0156\%$\\
$2.4$ & $0.369793-0.104941 i$ & $0.369449-0.105475 i$ & $0.165\%$\\
$2.8$ & $0.362609-0.107546 i$ & $0.362837-0.108064 i$ & $0.149\%$\\
$3.2$ & $0.352810-0.110103 i$ & $0.353895-0.110370 i$ & $0.302\%$\\
$3.6$ & $0.343773-0.110684 i$ & $0.345045-0.111689 i$ & $0.449\%$\\
$3.7$ & $0.341522-0.110857 i$ & $0.342767-0.111881 i$ & $0.449\%$\\
\hline
\hline
\end{tabular}
\caption{QNMs of the $\ell=3/2$ test scalar field ($s=0$) for the quantum corrected black hole ($M=1$) calculated using the JWKB formula at different orders and Padé approximants. The deviation is given in per cents.}\label{table5}
\end{table}

\begin{table}
\begin{tabular}{c c c c}
\hline
\hline
$\xi$ & WKB6Padé3 & WKB7Padé3 & rel. diff. \\
\hline
$0.4$ & $0.574082-0.096317 i$ & $0.574082-0.096317 i$ & $0.00004\%$\\
$0.8$ & $0.573899-0.096494 i$ & $0.573899-0.096497 i$ & $0.00041\%$\\
$1.2$ & $0.573099-0.097241 i$ & $0.573099-0.097242 i$ & $0.00014\%$\\
$1.6$ & $0.570937-0.099002 i$ & $0.570934-0.099001 i$ & $0.00053\%$\\
$2.0$ & $0.566542-0.101835 i$ & $0.566542-0.101835 i$ & $0\%$\\
$2.4$ & $0.559479-0.105134 i$ & $0.559467-0.105122 i$ & $0.00301\%$\\
$2.8$ & $0.550030-0.108126 i$ & $0.550029-0.108279 i$ & $0.0273\%$\\
$3.2$ & $0.538698-0.110280 i$ & $0.538830-0.110747 i$ & $0.0882\%$\\
$3.6$ & $0.526275-0.111913 i$ & $0.526658-0.112401 i$ & $0.115\%$\\
$3.7$ & $0.523111-0.112223 i$ & $0.523537-0.112690 i$ & $0.118\%$\\
\hline
\hline
\end{tabular}
\caption{QNMs of the $\ell=5/2$ test scalar field ($s=0$) for the quantum corrected black hole ($M=1$) calculated using the JWKB formula at different orders and Padé approximants. The deviation is given in per cents.}\label{table6}
\end{table}

\begin{table}
\begin{tabular}{c c c c}
\hline
\hline
$\mu$ & WKB6Padé3 & WKB10Padé5 & rel. diff.  \\
\hline
$0$ & $0.292931-0.097660 i$ & $0.292936-0.097660 i$ & $0.0015\%$\\
$0.04$ & $0.293647-0.097230 i$ & $0.293651-0.097230 i$ & $0.0014\%$\\
$0.08$ & $0.295796-0.095935 i$ & $0.295801-0.095935 i$ & $0.0014\%$\\
$0.12$ & $0.299389-0.093755 i$ & $0.299392-0.093755 i$ & $0.001\%$\\
$0.16$ & $0.304438-0.090657 i$ & $0.304438-0.090657 i$ & $0.0001\%$\\
$0.20$ & $0.310962-0.086591 i$ & $0.310957-0.086592 i$ & $0.0016\%$\\
$0.24$ & $0.318985-0.081487 i$ & $0.318963-0.081488 i$ & $0.0065\%$\\
$0.28$ & $0.328530-0.075273 i$ & $0.328467-0.075246 i$ & $0.0203\%$\\
$0.32$ & $0.339483-0.067839 i$ & $0.339461-0.067736 i$ & $0.0305\%$\\
$0.36$ & $0.351856-0.058897 i$ & $0.351866-0.058795 i$ & $0.0288\%$\\
$0.40$ & $0.365380-0.048789 i$ & $0.365478-0.048394 i$ & $0.111\%$\\
$0.44$ & $0.389911-0.037390 i$ & $0.380820-0.038668 i$ & $2.34\%$\\
$0.445$ & $0.395845-0.032388 i$ & $0.385450-0.038312 i$ & $3.01\%$\\
$0.447$ & $0.398303-0.029769 i$ & $0.388131-0.037754 i$ & $3.24\%$\\
\hline
\hline
\end{tabular}
\caption{QNMs of the $\ell=1$, $\xi =0.1$ test scalar field ($s=0$) for the quantum corrected black hole ($M=1$) calculated using the JWKB formula at different orders and Padé approximants. The deviation is given in per cents.} \label{table7}
\end{table}

\begin{table}
\begin{tabular}{c c c c}
\hline
\hline
$\mu$ & WKB6Padé3 & WKB7Padé3 & rel. diff.  \\
\hline
$0$ & $0.483643-0.096759 i$ & $0.483644-0.096759 i$ & $0.00013\%$\\
$0.04$ & $0.484148-0.096586 i$ & $0.484149-0.096586 i$ & $0.00013\%$\\
$0.08$ & $0.485665-0.096066 i$ & $0.485665-0.096066 i$ & $0.00013\%$\\
$0.12$ & $0.488196-0.095196 i$ & $0.488196-0.095196 i$ & $0.00013\%$\\
$0.16$ & $0.491747-0.093973 i$ & $0.491747-0.093972 i$ & $0.00013\%$\\
$0.20$ & $0.496326-0.092389 i$ & $0.496327-0.092389 i$ & $0.00011\%$\\
$0.24$ & $0.501944-0.090438 i$ & $0.501944-0.090437 i$ & $0.00009\%$\\
$0.28$ & $0.508613-0.088107 i$ & $0.508613-0.088107 i$ & $0.00004\%$\\
$0.32$ & $0.516349-0.085382 i$ & $0.516349-0.085382 i$ & $0.00003\%$\\
$0.36$ & $0.525171-0.082246 i$ & $0.525170-0.082246 i$ & $0.00016\%$\\
$0.40$ & $0.535102-0.078675 i$ & $0.535100-0.078675 i$ & $0.00040\%$\\
$0.44$ & $0.546166-0.074639 i$ & $0.546161-0.074639 i$ & $0.00094\%$\\
$0.48$ & $0.558390-0.070106 i$ & $0.558382-0.070099 i$ & $0.00182\%$\\
$0.52$ & $0.571793-0.065013 i$ & $0.571790-0.065004 i$ & $0.00171\%$\\
$0.56$ & $0.586415-0.059296 i$ & $0.586411-0.059289 i$ & $0.00136\%$\\
$0.60$ & $0.602265-0.052874 i$ & $0.602265-0.052874 i$ & $0.00003\%$\\
$0.64$ & $0.619312-0.045669 i$ & $0.619377-0.045677 i$ & $0.0105\%$\\
$0.68$ & $0.637710-0.038064 i$ & $0.637841-0.038128 i$ & $0.0228\%$\\
$0.72$ & $0.665429-0.024466 i$ & $0.655962-0.031741 i$ & $1.79\%$\\
%$0.73$ & $0.675415-0.012005 i$ & $0.668285-0.037198 i$ & $3.88\%$\\
%$0.732$ & $0.678173-0.008738 i$ & $0.674253-0.036928 i$ & $4.20\%$\\
\hline
\hline
\end{tabular}
\caption{QNMs of the $\ell=2$, $\xi =0.1$ test scalar field ($s=0$) for the quantum corrected black hole ($M=1$) calculated using the JWKB formula at different orders and Padé approximants. The deviation is given in per cents.}\label{table8}
\end{table}

\begin{figure*}
\resizebox{\linewidth}{!}{
\includegraphics{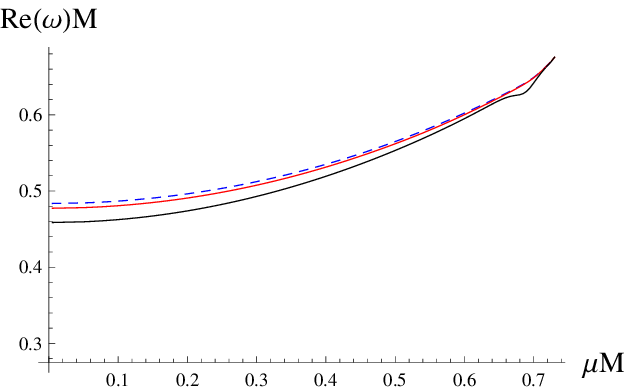}
\includegraphics{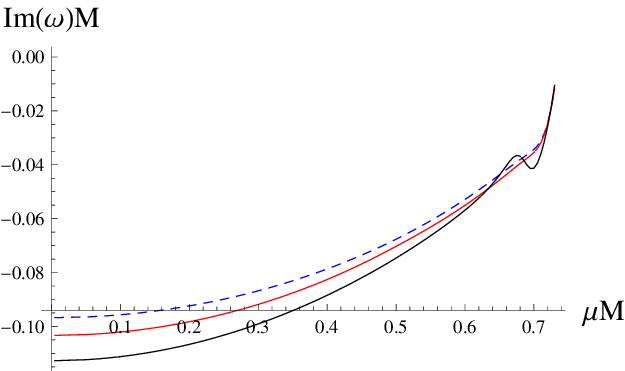}}
\caption{Real and imaginary part of the dominant ($n=0$) QNMs of the $\ell=2$ scalar field for the quantum corrected black hole calculated by the 6th order JWKB method with Padé approximants as functions of $\mu$: $\gamma=0.1$ (blue, dashed line), $\gamma=2$ (red, solid line), and $\gamma=3$ (black dots).}\label{fig:qnmsL2}
\end{figure*}

\begin{figure*}
\resizebox{\linewidth}{!}{
\includegraphics{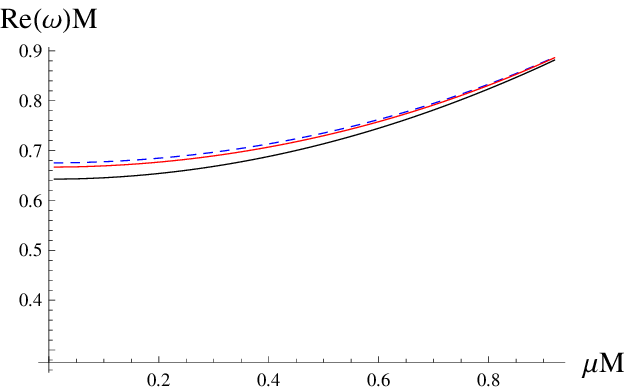}
\includegraphics{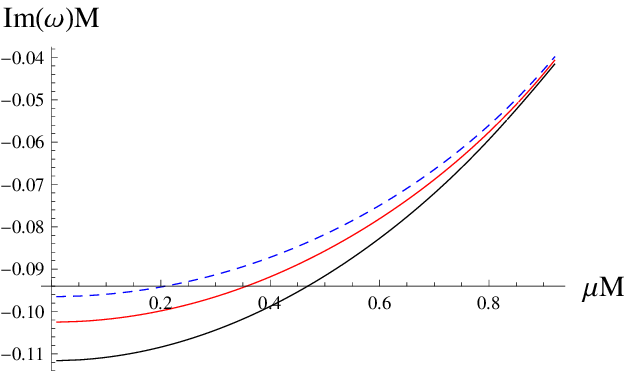}}
\caption{Real and imaginary part of the dominant ($n=0$) QNMs of the $\ell=3$ scalar field for the quantum corrected black hole calculated by the 6th order JWKB method with Padé approximants as functions of $\mu$: $\gamma=0.1$ (blue, dashed line), $\gamma=2$ (red, solid line), and $\gamma=3$ (black dots).}\label{fig:qnmsL3}
\end{figure*}

\begin{figure*}                            
\resizebox{\linewidth}{!}{\includegraphics{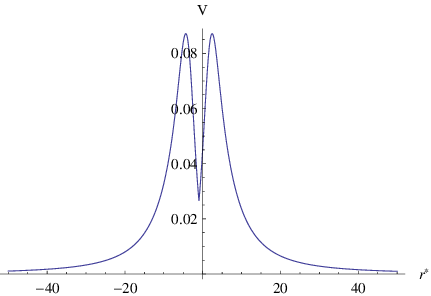}\includegraphics{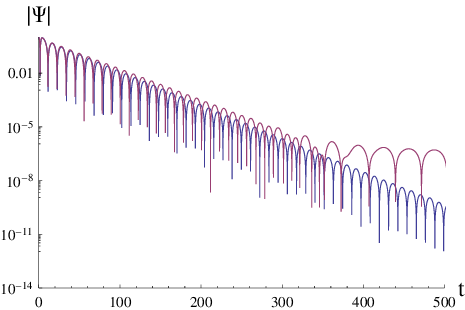}}          
\caption{Effective potential (left) and semi-logarithmic (right) time-domain profiles for wormhole perturbations: $\ell=1$, $\xi =5$, $M=1$, $\mu=0$ (blue) and $\mu=0.1$ (red). The Prony method gives $\omega=  0.27696 - 0.04321 i$ (for $\mu=0$) and  $\omega= 0.28212 - 0.03986 i$ (for $\mu=0.1$). For the massive perturbations the tails begin at about $t \sim 300$.}\label{fig:TDWH1}
\end{figure*}

\begin{figure*}                            
\resizebox{\linewidth}{!}{\includegraphics{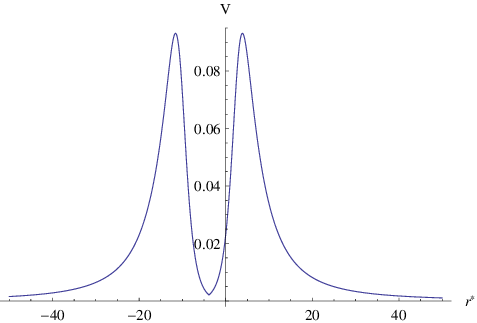}\includegraphics{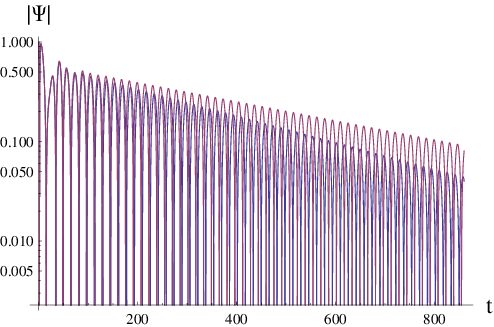}}          
\caption{Effective potential (left) and semi-logarithmic (right) time-domain profiles for wormhole perturbations near the transition: $\ell=1$, $\xi =4$, $M=1$, $\mu=0$ (blue) and $\mu=0.1$ (red). The Prony method gives $\omega=  0.19627 - 0.00309 i$ (for $\mu=0$) and  $\omega= 0.19888 - 0.00221 i$ (for $\mu=0.1$). For the massive perturbations the tails begin at about $t \sim 300$.}\label{fig:TDWH2}
\end{figure*}

\begin{figure}                            
\resizebox{\linewidth}{!}{\includegraphics{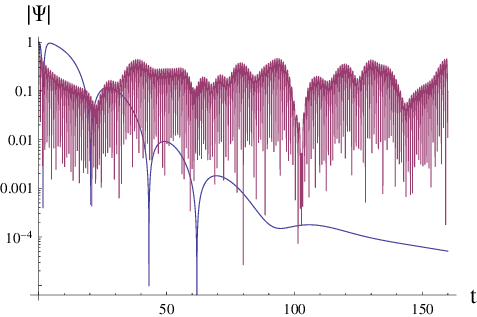}}          
\caption{Semi-logarithmic time-domain profiles for wormhole perturbations far from the transition: $\ell=0$, $\xi =6$, $M=1$, $\mu=0$ (blue) and $\mu=5$ (red). For large mass of the field the asymptotic tails start early and are modified by echoes owing to the presence of the second peak.}\label{fig:TDWH3}
\end{figure}

\vspace{3mm}
\section{QNMs and evolution of perturbations}\label{sec:qnms}

We begin with the most challenging case for both numerical and semi-analytic methods: the monopole mode with \( \ell = 0 \). The smaller the multipole number \( \ell \), the less accurate the JWKB approximation becomes~\cite{Konoplya:2019hlu}. Simultaneously, the time-domain profile for \( \ell = 0 \) typically exhibits only a few oscillations during the ringdown phase, making it difficult to extract quasinormal frequencies with high precision.

Despite this, the lowest multipoles are of particular interest, as they are often the least stable and can be responsible for the onset of instabilities (see, for example,~\cite{Bronnikov:1978mx}). However, there are also cases where instabilities arise at higher multipoles~\cite{Konoplya:2017lhs}.

Therefore, the poor agreement between the JWKB method and time-domain integration, as well as between different JWKB orders, for the lowest multipole, as illustrated in Table \ref{table1} and is not unexpected. Indeed, the JWKB values at different values and Padé approximants structure show not quite close results as shown in Table \ref{table1}. At the same time, as prescribed in \cite{Konoplya:2019hlu} the JWKB method can be trusted once different nearby JWKB orders produce close values of frequencies. Nevertheless, an important feature we observe in the $\ell=0$ time-domain profiles is the clear stability of the perturbations, which consistently decay into asymptotic tails.

In the case of massive field perturbations, the ringdown phase for \( \ell = 0 \) rapidly transitions into oscillatory tails with a power-law envelope~\cite{Koyama:2001qw,Moderski:2001tk,Rogatko:2007zz,Koyama:2001ee}. These tails effectively ``contaminate'' the ringdown signal, making it difficult to extract quasinormal frequencies with high accuracy.

For higher multipole numbers, however, the quasinormal frequencies can be determined with sufficient precision using both the WKB method and time-domain integration (see Figs. \ref{fig:TD1} and \ref{fig:TD2}). As shown in Tables \ref{table2}-\ref{table7}, the differences between values obtained using various (but close) WKB orders are minimal and significantly smaller than the observed physical effect — namely, the difference between the modes of a quantum-corrected black hole and its classical counterpart.

A central question when discussing QNMs of massive fields is whether \textit{arbitrarily long-lived modes}, known as quasi-resonances, can exist. In most cases, the answer is affirmative~\cite{Zhidenko:2006rs,Ohashi:2004wr}; however, there are notable exceptions in which quasi-resonances do not arise. These include the case of massive vector fields~\cite{Konoplya:2005hr}, massive scalar fields in certain braneworld scenarios~\cite{Zinhailo:2024jzt}, and asymptotically de Sitter spacetimes~\cite{Konoplya:2004wg,Lutfuoglu:2025hjy}. Thus, the existence of quasi-resonances must be examined on a case-by-case basis.

Strictly speaking, neither the WKB approximation nor time-domain integration provides sufficient precision to definitively identify the quasi-resonant regime. Nevertheless, as can be seen from Figs.~\ref{fig:qnmsL2} and~\ref{fig:qnmsL3}, as well as Tables~\ref{table7} and~\ref{table8}, the damping rate decreases significantly with increasing field mass \( \mu \). Extrapolating this trend suggests that the imaginary part of the frequency vanishes in the limit \( \mu \to \mu_c \), indicating the presence of quasi-resonances.

Even stronger evidence emerges when considering the behavior shown in Figs.~\ref{fig:qnmsL2} and~\ref{fig:qnmsL3}, where the quasinormal frequencies for different values of the coupling parameter \( \xi \) tend to merge as \( \mu \) increases. This indicates that, at larger \( \mu \), the spectrum becomes effectively independent of \( \xi \) and coincides with that of the Schwarzschild black hole. For the Schwarzschild case, the existence of quasi-resonances is well established~\cite{Ohashi:2004wr,Konoplya:2004wg}. Therefore, we conclude that there exists a critical mass \( \mu_c \) above which the damping rate of the fundamental quasinormal mode approaches zero. For \( \mu > \mu_c \), this mode disappears from the spectrum, and the first overtone takes over as the new fundamental mode.
 
The data in Tables~\ref{table4}–\ref{table6} show that the QNMs of neutrinos are also influenced by the quantum correction. Specifically, the damping rate increases monotonically with the coupling parameter \( \xi \), while the real part of the oscillation frequency decreases as \( \xi \) grows. It is evident that the impact of the quantum correction on the spectrum is at least an order of magnitude greater than the variation observed between different WKB orders, underscoring the significance of these corrections in the fermionic sector.

After the transition to the wormhole state, the effective potential develops a double-well structure with two symmetric peaks and a minimum at the throat. As a result, a very long-lived mode appears—even for massless fields—trapped within this potential structure by multipole reflections off the peaks (see Figs.~\ref{fig:TDWH1} and~\ref{fig:TDWH2}). When the mass of the field is introduced, the lifetime of this mode increases further, as evident from the same figures. Notably, near the threshold separating regular black holes from traversable wormholes, the QNMs become significantly longer lived compared to configurations far from the transition point.

In the regime \( \mu M \gg 1 \), the asymptotic tail sets in earlier, as can be observed in Fig.~\ref{fig:TDWH3}. The time-domain profiles in this case exhibit pronounced echo-like modulations at late times, characteristic of strong multiple scattering between the effective potential peaks. For massive fields, the signal becomes highly intricate and cannot be well-approximated by a simple analytic form. Nevertheless, the presence of slowly decaying, oscillatory tails is evident, underscoring the universality of this behavior for massive or effectively massive field dynamics near the black hole–wormhole threshold.

\section{Conclusions}
\label{sec:conclusions}

In this work, we investigated the QNMs of massive scalar and massless Dirac fields in the spacetime of regular black holes and traversable wormholes that arise within Covariant Effective Quantum Gravity. Using both the WKB approximation and time-domain integration, we have analyzed how quantum corrections affect the oscillation frequencies and damping rates of these perturbations.

We found that massive scalar fields exhibit slowly decaying, oscillatory tails and, in some cases, approach the regime of quasi-resonances—long-lived modes whose damping rate becomes arbitrarily small for  field mass larger than the critical. These quasi-resonances are especially prominent near the critical mass threshold, beyond which the fundamental mode disappears and higher overtones dominate. In addition, the fermionic (Dirac) sector is significantly affected by the quantum correction parameter \( \xi \), with damping rates increasing and oscillation frequencies decreasing as \( \xi \) grows.

Near the transition between a regular black hole and a traversable wormhole, we observed the emergence of echo-like features in the time-domain signal, particularly in the massive scalar case. Although the resulting waveforms become too complex to be fit by simple analytic forms, their qualitative features—such as the presence of long-lived, modulated tails—appear to be robust and characteristic of this quantum-corrected geometry.

Our analysis reinforces the importance of massive fields as probes of quantum gravity effects and highlights the utility of time-domain approaches in capturing complex dynamical behavior. Future work could focus on overtone structure \cite{Giesler:2019uxc,Silva:2024ffz,Stuchlik:2025ezz}, mode excitation due to different initial data, and extensions to rotating or higher-dimensional analogues of these geometries.

\vspace{4mm}
\begin{acknowledgments}
The author is grateful to Excellence Project PrF UHK 2205/2025-2026 for the financial support.
\end{acknowledgments}

\FloatBarrier
%\newpage
%\bibliographystyle{unsrt}
\bibliography{bibliography}
\end{document}